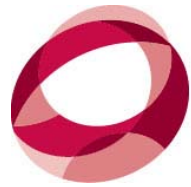

# Smart Wireless Communication is the Cornerstone of Smart Infrastructures

| Mary Ann Weitnauer | Jennifer Rexford | Nicholas Laneman |
| Georgia Institute of Technology | Princeton University | University of Notre Dame |

| Matthieu Bloch | Santiago Griljava | Catherine Ross |
| Georgia Institute of Technology | Georgia Institute of Technology | Georgia Institute of Technology |

Gee-Kung Chang
Georgia Institute of Technology

Summary

Emerging smart infrastructures, such as Smart City, Smart Grid, Smart Health, and Smart Transportation, need smart wireless connectivity. However, the requirements of these smart infrastructures cannot be met with today's wireless networks. A new wireless infrastructure is needed to meet unprecedented needs in terms of agility, reliability, security, scalability, and partnerships.

1. The Emergence of Smart Systems

We are at the beginning of a revolution in how we live with technology, resulting from a convergence of machine learning (ML), the Internet-of-Things (IoT), and robotics. A smart infrastructure monitors and processes a vast amount of data, collected from a dense and wide distribution of heterogeneous sensors (e.g., the IoT), as well as from web applications like social media. In real time, using machine learning, patterns and relationships in the data over space, time, and application can be detected and predictions can be made; on the basis of these, resources can be managed, decisions can be made, and devices can be actuated to optimize metrics, such as cost, health, safety, and convenience.

2. Wireless is the Cornerstone of Future Smart Systems

Smart Infrastructure cannot exist without wireless communication. Mobile access to the infrastructure by people, machines, and sensors drives the need for radio connections. Even for static embedded sensors, installing cable to connect the sensors to the internet is often prohibitively expensive; in contrast, wireless connection lowers the capital expenditure of sensor deployment. Asset tracking, a huge market in IoT, requires a wireless connection between the infrastructure and the object being tracked. The proliferation of smart mobile devices, with their high flexibility, versatility, and computational capacity, has necessitated wireless wideband networks with ubiquitous coverage. In 5G, the extremely high data rates per unit area will be supplied in large part by dense deployments of small cells, which may be connected to a Baseband Server (BBS) via a directional wireless link, to reduce cost of deployment. Thus, a smart phone or sensor signal may traverse two wireless hops before hitting optical fiber.

The developing wireless technologies are shaping future opportunities in smart transportation, which is the largest application of IoT in cities. Autonomous and connected vehicles and management of traffic flow among other IoT applications will lead to better economic value, safety, efficiency, security and sustainability. For example, better tracking and control of public transportation enabled by a future wireless infrastructure has the potential to reduce the up to 70% of "buffer time" commuters build into their travel times, leading to savings of $60 billion annually.[1]

---

[1] McKinsey Global Institute, "The internet of Things: Mapping the Value Beyond Hype", 2015.

Real-time data, centralized traffic control systems, integration of pedestrian and bicycle infrastructure, smart metering for parking and utilities, travel time savings, reduced fuel consumption, reallocation of land uses, enhanced car sharing, management of water, sewer and electricity and reduced transport system failures are among a few of the benefits of investment in wireless technology.

3. Needs in Advanced Wireless Technologies

In spite of the quality of today's wireless networks, there remain very difficult challenges to meet the needs of the emerging smart infrastructure, specifically in terms of number, bandwidth, and heterogeneity of connections, coverage, latency, support for low-energy devices, privacy, security, support for some measure of free public access, and resilience.

**Scaling to a large number of devices**: The number of radio connections will see staggering growth in the next decade. The number of devices connected to the Internet surpassed the number of people on Earth (6.7B) in 2008[2], and is expected to be about 50B by 2020.[3] These devices will create an immense amount of data that far exceeds the capacity of our current network. For example, one driverless vehicle will generate about one gigabyte per second.[4] The U.S. Smart Grid is expected to source about 1000 petabytes per year, which is five times what AT&T's entire network carried in 2010.[5]

**Scaling mobile data volume by 1000X**: Worldwide mobile data traffic grew 63% in 2016 and is expected to account for 66% of internet traffic by 2020.[6] The data rate need per device is expected to increase by a factor of 100, owing to emerging applications such as HD video streaming, virtual reality, and 3D gaming. Several new air interface technologies, such as millimeter wave (mmWave), massive MIMO, and small cell densification, are being developed to support the anticipated 1000X increased mobile data load.[7] While these technologies offer much higher data rates in static scenarios, they are not as robust to mobility as current microwave (e.g., mobile and WiFi) networks. For example, small cells induce more frequent handovers and the mmWave channel will be much more sensitive to terminal motion and blockage of the line-of-sight by objects in the channel. An existing technology that was intended to scale mobile data volume, the Cloud Radio Access Network (CRAN), performs in cloud data centers the signal processing functions that were traditionally performed in base stations. CRANs offer extreme improvements in operating cost and energy efficiency as the signal processing base band units (BBUs) in the data center can be optimally and dynamically reallocated to process the data from the low-complexity remote radio heads (RRHs) as traffic loads shift spatially over time. Also, the signals from several RRHs can be jointly processed in the data center to achieve high spectral efficiency from distributed MIMO processing.[8] However, the mobile fronthaul network, which connects the wireless edge to the CRAN data center, has been shown to be insufficient in terms of data capacity and latency and is not financially feasible for an operator unless the operator already has the fiber.[9]

---

[2] Nicholas Jackson, "Infographic: The Internet of Things," The Atlantic, Jul 22, 2011, https://www.theatlantic.com/technology/archive/2011/07/infographic-the-internet-of-things/242073/, downloaded 5-14-2017.
[3] Jeff McCracken "Unlocking the power of analytics with the right data in the right place," Metering and Smart Energy International, May 10, 2017, https://www.metering.com/magazine_articles/grid-data-analytics-jeff-mccracken/, downloaded 5-14-2017.
[4] L. Mearian. Self-Driving Cars Could Create 1GB of Data a Second. Accessed on Apr. 7, 2016. [Online]. Available: http://www.computerworld.com/article/2484219/emerging-technology/ self-driving-cars-could-create-1gb-of-data-a-second.html
[5] N. Cochrane. (Mar. 23, 2010). US Smart Grid to Generate 1000 Petabytes of Data a Year. Accessed on Apr. 7, 2016. [Online]. Available: http://www.itnews.com.au/news/us-smart-grid-to-generate1000-petabytes-of-data-a-year-170290#ixzz458VaITi6
[6] White paper: Cisco VNI Forecast and Methodology, 2015-2020, 2016, http://www.cisco.com/c/en/us/solutions/collateral/service-provider/visual-networking-index-vni/complete-white-paper-c11-481360.pdf , downloaded 5-20-2017.
[7] Ian F. Akyildiz, Shuai Nie, Shih-Chun Lin, Manoj Chandrasekara, "5G roadmap:10 key enabling technologies," Computer Networks, Vol. 106, pp. 17–48, 2016.
[8] Chih-Lin, I., Jinri Huang, Ran Duan, Chunfeng Cui, Jesse Xiaogen Jiang, and Lei Li. "Recent progress on C-RAN centralization and cloudification." IEEE Access 2 (2014): 1030-1039.

**Low latency for real-time control**: Latency, or the round-trip delay in a network, must decrease by more than a factor of 10 to support real-time control applications such as Tactile Internet, multi-player gaming, and virtual reality.[10] Applications involving real-time interaction with the user need about 1ms latency or else the user experience is degraded. Machine type communications, such as vehicle-to-vehicle communications, also demand about 1ms latency. However, in today's LTE mobile communications network, the latency in the data plane is about 15ms.[11]

**Supporting energy-efficient devices**: Besides the high-functioning and bandwidth-hungry smartphones, tablets, and laptops, the network of the future must support vast numbers of moderate-to-low data rate and energy-constrained IoT devices, e.g., sensors that are powered by harvesting ambient energy or whose batteries need to last for many years, to generate the data that feeds the smart infrastructure. The current mobile communications network is not optimized for these types or numbers of IoT devices.

**Protecting security and privacy**: As more devices are connected to each other and to the internet, our wireless infrastructure must be protected from cyberattacks, including new threats on (and from) IoT devices. The proliferation of IoT also raises tremendous privacy challenges, since these sensors can monitor human location, activity, and even mood. These devices are created by a wide range of companies for a dizzying array of purposes, making it difficult to apply existing security and privacy techniques designed for desktop or smart phone systems.

**Integrated free access option**: The trends in free WiFi access tell us that a new integrated wireless infrastructure must support some amount of free access. While the free municipal WiFi networks of the mid 2000's were too expensive and were shut down after a few years, the number of businesses that offer free WiFi has grown.[12] A 2014 study found that businesses that offered free Wi-Fi attracted more customers and those customers spent more time and more money on the premises.[13] In 2013, more than 90% of the largest 150 U.S. airports offered free WiFi, and those that did not suffered a competitive disadvantage.[14]

**Meeting the needs of the community**: Designing a new wireless infrastructure that truly meets the needs of communities is extremely challenging and demands a vision beyond what the wireless industry alone can provide. More partnerships with cities and civic organizations are needed to develop new technology that will have the intended benefits rather than creating new problems.[15] It is easy for technologists, engineers, and researchers to dream up high tech "solutions," but much harder for these professionals to work with local authorities and organizations, such as law enforcement, public housing, transportation, schools, government, and philanthropic organizations, to find solutions that will raise the human condition, such as broadening access in rural and low

---

[9] Mugen Peng, Yaohua Sun, Xuelong Li, Zhendong Mao, and Chonggang Wang, "Recent advances in cloud radio access network: system architectures, key techniques, and open issues," IEEE Communication Surveys & Tutorials, Vol. 18, No. 3, Third Quarter, 2016.

[10] G. Fettweis, The tactile Internet: applications and challenges, IEEE Veh. Technol. Mag. 9 (1) (2014) 64–70.

[11] 3GPP TS 25.913, Requirements for evolved UTRA and evolved UTRAN, 2009

[12] Whatever happened to municipal Wi-Fi? The Economist, July 26, 2013, http://www.economist.com/blogs/babbage/2013/07/wireless-networks

[13] "The Benefits of Offering Free Customer Wi-Fi," Inc.com, Apr 22, 2015, https://www.inc.com/comcast/the-benefits-of-offering-free-customer-wifi.html

[14] Alex Konrad, "Who's paying for your free airport wifi?" Feb 10, 2014, https://www.forbes.com/sites/alexkonrad/2014/01/22/airport-wifi-free/#235ca52c281a

[15] Nancy Rodriguez Ph.D., Chris Tillery, "From the Directors: Decision Factors for Adopting a Technology," The Police Chief Magazine, October 2016. http://www.policechiefmagazine.org/wp-content/uploads/PoliceChief_October_2016LORES1.pdf downloaded 4-28-2017

income areas and thereby reducing the "Digital Divide."

In summary, we need a wireless infrastructure that is ubiquitous (across a large and diverse country), performant, reliable, resilient and secure (worthy of society's trust), energy-efficient, and cost-effective, to realize smart infrastructures that can support the most advanced 21$^{st}$ century society.

4. Recommendations

To meet the demands of smart infrastructures, the wireless infrastructure of the future must meet unprecedented demands in multiple dimensions: integration, high bandwidth and agility, reliability, security, and partnerships. Specifically, we recommend enhanced investment in the following key technology areas.

FLEXIBLE INFRASTRUCTURE
- Software defined networking, including software defined radio, to create a flexible infrastructure that can adapt to changing conditions to offer good performance, reliability, and security in dynamic environments, and handle tasks that heterogeneous, low-power IoT devices are not capable of performing themselves.
- Application of machine learning and big data analysis for dynamically analyzing wireless signals and performance monitoring, simultaneously in multiple radio bands and air interfaces, to optimize wireless channels and network configuration (e.g., power, channel assignment, beam direction, protocol parameters, etc.) for better network performance, especially in regions with dense deployments.
- Optimum distribution of computation, storage, *and* communication resources *ranging from* the cloud to the mobile edge (e.g., fog computing[16]), for better performance (e.g., lower latency and higher reliability) and cost (e.g., lower bandwidth usage on the wired network to the cloud). This includes better design of (i) the fiber-wireless interface to optimize utilization of the fronthaul optical bandwidth, (ii) the high-speed switches to facilitate fog computing and dynamic clustering of distributed BBUs.
- Virtualization of the wireless networks to decouple services from ownership of the wireless network infrastructure, for greater flexibility and cost-effectiveness, and for security (to provide isolation for critical traffic for smart infrastructure).
- *Integration* of limited public access options for businesses that wish to offer free access.
  For example, knowing in the beginning that a free public access channel will eventually be offered almost ubiquitously by businesses, include this service option as part of the design of the new integrated architecture.

HIGH-BANDWIDTH AND AGILE WIRELESS TECHNOLOGIES
- Dynamic clustering of communication resources along the edge, achieved with high speed, extreme bandwidth routers, to support low-latency applications and to reduce the traffic on the fronthaul network. The resources should include distributed BBUs and RF bands in dynamic spectrum aggregation and management.
- Study of extremely high bandwidth radio access technologies, such as ultra-dense cells, millimeter wave, TeraHerz band, massive MIMO, and carrier aggregation, in practical mobile environments and in the context of the software defined network, especially considering the impact of the fragility of these wireless links on the fiber-wireless interface and on dynamic BBU clustering and fog computing.
- Effective techniques such as caching and mobile infrastructure for handling intermittent connectivity to work with low-power devices that may be temporarily unreachable to save energy.

---

[16] Mung Chiang and Tao Zhang, "Fog and IoT: an overview of research opportunities," IEEE Internet of Things Journal, Vol. 3, No. 6, December 2016.

RELIABILITY, SECURITY, AND RESILIENCE
- Security solutions for authentication, preventing wireless infrastructure attacks and privacy leaks, IoT security/privacy, etc. need for end-to-end security solutions that work for users who are (i) moving from one location to another and (ii) using different access network technologies across time, or even at the same time.
- New programming tools and software stacks to lower the barrier for creating effective wireless services that have good performance, reliability, energy efficiency, etc.
- Energy-efficient wireless devices (especially sensors and actuators) that can collect and report data and effect change in the physical world (for IoT, smart infrastructure, etc.), as well as techniques for energy harvesting.
- Seamless integration of heterogeneous access technologies (WiFi, cellular, wired, etc.), including using multiple at the same time, for better coverage, performance, and reliability.
- A usable failure mode that provides low bandwidth reliable communication in disaster scenarios.

PARTNERSHIPS

- Research projects done in partnership with local authorities and organizations, such as law enforcement, public housing, transportation, schools, government, and philanthropic organizations. The projects should have different scopes to enable a diverse array of participants, ranging from small pilot programs with low entry barriers, to city-scale projects, such as NSF PAWR, to accelerate technology and ecosystem innovations.
- Partnerships with NIST and FCC to experiment with dynamic spectrum and unlicensed spectrum, and potentially use these more aggressively going forward, to support research and innovation.
- Partnerships with philanthropic organizations, such as the Rockefeller 100 Resilient Cities project, to help cities deal with shocks and stressors.
- Partnerships with standards organizations to create open-source software and open interfaces/standards for programmable wireless networks, manage wireless infrastructure, and create wireless applications to enable innovation and interoperability.
- Partnerships with academia in the training of students, government employees (e.g., law enforcement), and retraining of workers in many fields, to create and use wireless technologies to build solutions to address important societal and business problems.

*This white paper is part of a series of white papers on Intelligent Infrastructure, written by community members. The full series of papers can be found at this website- http://cra.org/ccc/resources/ccc-led-whitepapers/#infrastructure. The Intelligent Infrastructure white paper series was a collaboration between the Computing Community Consortium and the Electrical and Computer Engineering Department Heads Association (ECEDHA).*

*This material is based upon work supported by the National Science Foundation under Grant No. 1136993. Any opinions, findings, and conclusions or recommendations expressed in this material are those of the authors and do not necessarily reflect the views of the National Science Foundation.*

*For citation use: Weitnauer M., Rexford J., Laneman N., Bloch M., Griljava S., Ross C., & Chang G. (2017) Smart Wireless Communication is the Cornerstone of Smart Infrastructures. http://cra.org/ccc/resources/ccc-led-whitepapers/*